\documentclass[a4paper,twoside,10pt]{article}

\input{jgrg21.sty}
\begin{document}
%
\pagestyle{fancy}
\fancyhead{}
  \fancyhead[RO,LE]{\thepage}
  \fancyhead[LO]{K. Shiraishi}                  
  \fancyhead[RE]{Graph-theory induced gravity}
\rfoot{}
\cfoot{}
\lfoot{}
\label{P20}    
\title{%
Einstein Universe under Deconstruction: the case with degenerate fermions 
}
%
\author{%
  Nahomi Kan\footnote{Email address: kan@yamaguchi-jc.ac.jp}$^{(a)}$,
  Koichiro Kobayashi\footnote{Email address: m004wa@yamaguchi-u.ac.jp}$^{(b)}$,
  and
  \underline{Kiyoshi Shiraishi}\footnote{Email address:
shiraish@yamaguchi-u.ac.jp}$^{(b)}$ }
%
\address{%
  $^{(a)}$Yamaguchi Junior College, Hofu-shi, Yamaguchi 747--1232, Japan\\
  $^{(b)}$Yamaguchi University, Yamaguchi-shi, Yamaguchi 753--8512, Japan\\}
%
\abstract{
We study self-consistent static solutions for an Einstein universe in a graph-based
induced gravity. In the generalization of the deconstruction model based on the graph, the
eigenvalues of the graph Laplacian and the adjacent matrix gives the mass spectrum the
particles. Thus we can easily control UV divergences at one-loop level in such a model. We
use the calculation method with the spectrum distribution function of the graph and search
for the static solution supported by the degenerate pressure of the fermion (at zero
temperature). The report is based on 
{\tt arXiv:1110.5697}. 
}

\section{Introduction}
In our previous work \cite{P20_ref1}, the induced gravity \cite{P20_ref2} model without UV
divergences at one-loop level has been constructed by a generalized method of Dimensional
Deconstruction (DD) \cite{P20_ref3} and a self-consistent solution for an Einstein static
universe has been obtained.

In this brief report, we show the existence of a
self-consistent Einstein universe in which strongly degenerate fermions by the calculation
method using the spectral density function of graphs.

\section{Induced gravity}
Induced gravity  has been studied by many authors \cite{P20_ref2}.
The one-loop effective
action can systematically be expressed by an integral form using
Schwinger's proper time method as
\begin{equation}
\frac{1}{2}{\rm Tr\,}\ln H=-\frac{1}{2}\int_0^\infty
\frac{dt}{t}{\rm Tr\,}\left[e^{-tH}\right]\,,
\end{equation}
where $H$ is a Hessian operator which appears in the free-field action of a 
matter field.
The expansion in terms of the Seeley-DeWitt coefficients
can be written as
\begin{equation}
{\rm Tr\,}\left[e^{-tH}\right]=\frac{1}{(4\pi t)^2}\int
d^4x\sqrt{|\det g_{\mu\nu}|}\left[{\rm Tr\,} a_0+t\,{\rm Tr\,} a_1+t^2\,{\rm Tr\,}
a_2+o(t^3)\right]\,,
\end{equation}
where 
$g_{\mu\nu}$ denotes the spacetime metric
and ${\rm Tr\,}$ means the trace over the spacetime indices.  The one-loop
effective action for the background fields is given by the collection of the contribution
of various matter fields.

The UV divergences arise from the integration in the
vicinity of $t=0$.
If we introduce a UV-cutoff scale $\Lambda$,
the lower bound of the integration on $t$ is replaced to
${1/\Lambda^2}$.
The divergent parts in terms of the cut-off $\Lambda$ are
\begin{equation}
\frac{1}{64\pi^2}(N_0-2N_{1/2}+2N_1)\Lambda^{4}\,
\quad {\rm and} \quad
\frac{1}{192\pi^2}(N_0+N_{1/2}-4N_1)\Lambda^2 R\,,
\end{equation}
where $N_0$ is the number of minimal scalar degrees of freedom,
$N_{1/2}$ is the number of two-component fermion fields,
$N_1$ is the number of massless vector fields, and $R$ is the scalar curvature
constructed from the metric $g_{\mu\nu}$.

The conditions for their cancelations are solved by
$N_0=2N$, $N_{1/2}=2N$, and $N_1=N$, where $N=1, 2, 3,\dots$.

For massive fields, since
\begin{equation}
\sum_{i=1}^{N_s}e^{-(m^2_s)_i
t}=N_s-t\,\sum_{i=1}^{N_s}(m^2_s)_i+t^2\,\frac{1}{2}\sum_{i=1}^{N_s}
(m^4_s)_i+\cdots\equiv
N_s-t\,{\rm Tr\,} M^2_s+t^2\,\frac{1}{2}{\rm Tr\,} M^4_s+\cdots\,,
\end{equation}
(where $M^2_s$ is the mass-squared matrix for spin-$s$ field),
the condition on mass-squared matrix for the cancelation of divergences should be
$
{\rm Tr\,} M^2_S-4\,{\rm Tr\,} M^2_{D}+3\,{\rm Tr\,} M^2_V=0
$,
where $M_S^2$ is the mass-squared matrix for the scalar fields,
$M_D^2$ is that for the Dirac fields, and
$M_V^2$ is that for the vector fields.

Now, we construct the field theories with mass matrices which satisfy the  cancelation
conditions.

\section{Graph and mass matrices}
We remember
the concept of DD \cite{P20_ref3}.
A moose diagram is used to
describe this theory, and is no more than a graph. 
The $N$-sided polygon is identified as an example of simple graphs, a
cycle graph $C_N$. 
A graph $G$ consists of a vertex set ${\cal V}$
and  an edge set ${\cal E}$,
where an edge is a pair of distinct vertices of $G$.
The
graph with directed edges is dubbed as a directed graph. An oriented edge
$e=[u,v]$  connects the 
origin $u=o(e)$ and the terminus $v=t(e)$.

Now we introduce several matrices that are naturally associated with a
graph \cite{P20_ref4,P20_ref5}. They are
the incidence matrix $E(G)=(E)_{ve}$, 
the adjacency matrix $A(G)=(A)_{vv^{\prime}}$,
the degree matrix $D(G)=(D)_{vv^{\prime}}$, and
the graph Laplacian (or combinatorial Laplacian) $\Delta (G)=(\Delta)_{vv^{\prime}}$.
The  relations among them are $\Delta=D-A$, and $\Delta= EE^T$.
The important identities are ${\rm Tr}\,A=0$, and ${\rm Tr}\,A^2={\rm Tr}\,D$.
Thus the relations ${\rm Tr}\,\Delta={\rm Tr}\,D$ and
${\rm Tr}\,\Delta^2={\rm Tr}\,D^2+{\rm Tr}\,D$ hold.

The model of vector fields, whose mass-squared matrix is $f^2\Delta$, is
\cite{P20_ref5}
\begin{equation}
{\cal L}_V=-\frac{1}{4}\sum_{v\in {\cal V}}F^v_{\mu\nu}F_v^{\mu\nu}-
\sum_{e\in {\cal E}}({\cal D}_\mu U_e)^\dagger ({\cal D}^\mu U_e)\,,
\end{equation} 
where the covariant derivative is
${\cal D}^\mu U_e\equiv (\partial^\mu+iA^\mu_{t(e)}-A^\mu_{o(e)})U_e$
 with $|U_e|=f$. Here $f$
is a constant with the dimension of mass. 
Similarly, any kind of fields can be associated with a graph
and their mass-squared matrix can be written by using
the graph Laplacian.
For scalar fields, we assign a scalar field
$\phi_v$ to each
vertex $v$ of
$G$.
A mass term for scalar fields can be constructed as
$f^2\sum_{v,v'\in {\cal V}}\phi_v\Delta_{vv'}
\phi_{v'}$.
For spinor fields, the mass term can be expressed by using
the incidence matrix
$E$. 
For example, the Lagrangian density of fermion fields can be written
as \cite{P20_ref5}
\begin{equation}
-\sum_{v\in {\cal V}}\bar{\psi}_{Rv}{\it
D\!\!\!\!/~}\psi_{Rv}-\sum_{e\in {\cal E}}\bar{\psi}_{Le}{\it
D\!\!\!\!/~}\psi_{Le}-f\sum_{e\in
{\cal E}}\sum_{v\in {\cal V}}[(\bar{\psi}_{Le}(E^T)_{ev}\psi_{Rv}+h.
c.]\,,
\label{P20_label_2}
\end{equation}
where the subscripts $L$ and $R$ denote left-handed and right-handed
fermions, respectively.
Namely, the left-handed fermions are assigned to the edges while the right-handed
ones are assigned to the vertices. The mass spectrum of
fermions governed by the Lagrangian (\ref{P20_label_2}) is also given by
the eigenvalues of the graph Laplacian \cite{P20_ref5}.
                          
Therefore, the UV divergences can be controlled using the graph Laplacian
and we can construct the models of UV-finite induced gravity.
We prepare three graphs, $G_S$, $G_D$ and $G_V$.
All these graphs have $N$ vertices.
If the graphs associated to their field have the same degree matrix, we find
\cite{P20_ref5}
\begin{equation}
{\rm Tr\,} M_S^2={\rm Tr\,} M_D^2={\rm Tr\,} M_V^2\,\quad {\rm and}\quad
{\rm Tr\,} M_S^4={\rm Tr\,} M_D^4={\rm Tr\,} M_V^4\,.
\end{equation}
Therefore we find that the induced vacuum energy
and the inverse of the Newton constant at one-loop can be calculated
for selected graphs
\cite{P20_ref7}.
Suppose that we select a type
of non-simple graphs
$G_{\{n_i\}}=C_{n_1}\cup C_{n_2}\cup
\cdots$, which has $N$ vertices.
Then we can choose different sets $\{n_i\}$ for scalar, Dirac, and vector fields
in a model in order to obtain non-zero value for the Newton and cosmological
constants \cite{P20_ref7}.

\section{The effective action in an Einstein universe}
We assume that the background geometry is given by a static Einstein universe.
Carrying out the integration over $t$,
we expand the effective action in terms of the modes of Laplacian on $S^3$ 
(with the radius
$a$). 
The regularized mode sum for a scalar field with mass $m$ is found to be
\begin{equation}
\Sigma'_S(m^2a^2)=\sum_{\ell=1}^\infty \ell^2\left[
\sqrt{\ell^2+m^2a^2-1}-
\ell \left(1+\frac{m^2a^2-1}{2\ell^2}-
\frac{(m^2a^2-1)^2}{8\ell^4}\right)\right]
-\frac{5m^2a^2-6}{120}-\frac{1}{8}\gamma\,,
\end{equation}
(where  $\gamma$ is the Euler-Mascheroni
constant),
and similar expressions are obtained for the fermion field and the vector field.

Using these expressions, we obtain the effective action
\begin{equation}
\frac{1}{2a}
\sum_i\left[\Sigma'_S((m_0^2)_ia^2)-
\Sigma'_D((m_{1/2}^2)_ia^2)+\Sigma'_V((m_{1}^2)_ia^2)\right]\,.
\end{equation}

\section{Spectral density function of a graph}
For $C_N$,
eigenvalues of the adjacency matrix are
$
\lambda_k=2\cos\frac{2\pi k}{N}\,,\quad (k=0, 1, \dots, N-1)\,.
$
For a large $N$, we find
\begin{equation}
\lim_{N\rightarrow\infty}\frac{1}{N}
\sum_{k=0}^{N-1}f\left(\lambda_k\right)=
\int_0^1f(2\cos \pi
t)dt=\frac{1}{\pi}\int_{-2}^{2}f(x)\frac{dx}{\sqrt{4-x^2}}\,,
\end{equation}
then we can define  
the spectral density function of the cycle graph \cite{P20_ref8} as
\begin{equation}
\rho_\infty(x)=\left\{
\begin{array}{cc}
\frac{1}{\pi}\frac{1}{\sqrt{4-x^2}}\,, & {\rm for~}-2<x<2\\
0\,, & {\rm others}
\end{array}
\right.\qquad {\rm for~cycle~graphs}\,.
\end{equation}

A large $N$ means that  the part of
the effective action for a small $a$ (the radius of the universe) is dominant.
Then we approximate the effective action for a small $a$ as
\begin{equation}
\Omega_0(fa)\equiv\frac{1}{2a}\int_{-2}^2\left[\Sigma'_S(f^2a^2(2-x))-
\Sigma'_D(f^2a^2(2-x))+\Sigma'_V(f^2a^2(2-x))\right]
\frac{N}{\pi\sqrt{4-x^2}}dx\,.
\end{equation}
The graphs of the type of $C_n\cup C_m\cup\cdots$ have the same spectral density function
for a large
$N$.


\section{Strongly degenerate fermions} 

The thermodynamical potential with strongly degenarate fermions ($T=0$) can be 
expressed only by the chemical
potential
$\mu$ and the mass spectrum of fermions.

Applying the spectral density function to this, we get
\begin{eqnarray}
&&\Omega_D=-\frac{2\pi^2
a^3}{12\pi^2}\int_{-2}^2\theta(\mu-m(x))\nonumber \\  &&\times
\left[\mu\sqrt{\mu^2-m^2(x)}\left(\mu^2-\frac{5}{2}m^2(x)\right)+
\frac{3}{2}m^4(x)\ln\left(\frac{\mu}{m(x)}+
\sqrt{\frac{\mu^2}{m^2(x)}-1}\right)\right]
\frac{N}{\pi\sqrt{4-x^2}} dx\,,
\end{eqnarray}
where $m^2(x)\equiv f^2(2-x)$ and  $\theta(y)$ is the step
function.
This is the main contribution for a large $N$.

The Einstein equations can be written by using the thermodynamical potential $\Omega$,
which includes the vacuum contribution:
\begin{equation}
\frac{\partial(\mu^{-1}\Omega)}{\partial(\mu^{-1})}=0\,
\quad {\rm and}
\quad\frac{\partial(\mu^{-1}\Omega)}{\partial a}=0\,.
\end{equation}

\begin{figure*}[h]
\centering
\includegraphics[keepaspectratio=true,width=7cm]
{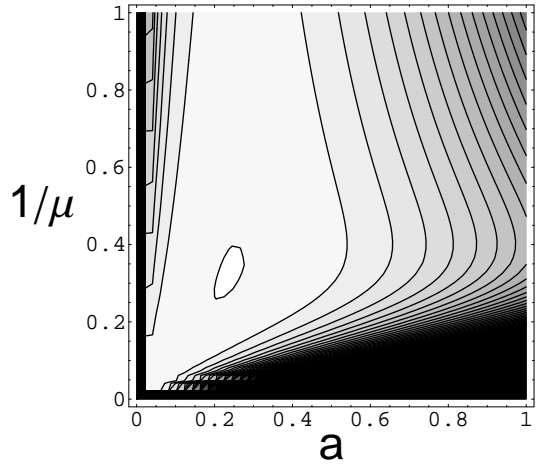}
\caption{%
A contour plot of $\frac{1}{N}\mu^{-1}\Omega$.
}
\label{P20_fig1.eps}
\end{figure*}

In Figure~\ref{P20_fig1.eps}, we show the contour plots for $\Omega/\mu$ obtained by
numerical calculations, whose extremum provides a self-consistent
solution. The horizontal axis indicates the scale factor
$a$, while the vertical one $1/\mu$, in the unit of $f$.
One  (unstable) solution for a self-consistent Einstein universe can be found. 
The Casimir
effects is essential in this case.

\section{Summary and prospects}
We have shown the construction of Graph-based (calculable) induced gravity models.
For a large $N$ (the number of fields), 
a small $a$ (the radius of the universe),
the effective potential (mainly the Casimir energy) and
the thermodynamical potential for the degenerate fermions are evaluated by using the
spectral density function of graphs.
We found the existence of a self-consistent solution for a static Einstein universe.

In future work, the trace formula for graph spectrum will be directly applied to the
one-loop calculations.



\begin{thebibliography}{99}

\bibitem{P20_ref1}
N.~Kan and K.~Shiraishi, Prog. Theor. Phys. {\bf 121} (2009) 1035.

\bibitem{P20_ref2} 
For a review, M.~Visser, Mod. Phys. Lett. {\bf A17} (2002) 977.

\bibitem{P20_ref3}
N.~Arkani-Hamed, A.~G. Cohen and ~H.
Georgi, Phys. Rev. Lett. {\bf 86} (2001) 4757;
C.~T.~Hill, S.~Pokorski and J.~Wang, Phys. Rev. {\bf D64} (2001) 105005.

\bibitem{P20_ref4}
B.~Mohar, ``The Laplacian spectrum of graphs'', in 
{\it Graph Theory, Combinatorics, and Applications},
ed. Y.~Alavi {\it et al.} (Wiley, New York, 1991), p. 871;
R.~Merris, Linear Algebra Appl. {\bf 197} (1994) 143.

\bibitem{P20_ref5}
N.~Kan and K.~Shiraishi, J. Math. Phys. {\bf 46} (2005) 112301.

\bibitem{P20_ref7}
N.~Kan and K.~Shiraishi, Prog. Theor. Phys. {\bf 111} (2004) 745.

\bibitem{P20_ref8}
A.~Hora and N.~Obata,
{\it Quantum Probability and Spectral Analysis of Graphs},
Springer, Berlin Heidelberg, 2007. 

\end{thebibliography}
\end{document}